# Quantitative DMS mapping for automated RNA secondary structure inference

Pablo Cordero[1], Wipapat Kladwang[2], Christopher C. VanLang[3], and Rhiju Das[1,2,4*]

Departments of Biomedical Informatics[1], Biochemistry[2], Chemical Engineering[3], and Physics[4], Stanford University, Stanford CA 94305

Supporting Information Placeholder

**ABSTRACT:** For decades, dimethyl sulfate (DMS) mapping has informed manual modeling of RNA structure *in vitro* and *in vivo*. Here, we incorporate DMS data into automated secondary structure inference using a pseudo-energy framework developed for 2´-OH acylation (SHAPE) mapping. On six non-coding RNAs with crystallographic models, DMS-guided modeling achieves overall false negative and false discovery rates of 9.5% and 11.6%, comparable or better than SHAPE-guided modeling; and non-parametric bootstrapping provides straightforward confidence estimates. Integrating DMS/SHAPE data and including CMCT reactivities give small additional improvements. These results establish DMS mapping – an already routine technique – as a quantitative tool for unbiased RNA structure modeling.

Understanding the many biological functions of RNAs, from genetic regulation to catalysis, requires accurate portraits of the RNAs' folds. Among biochemical tools available for interrogating RNA structure, chemical mapping or "footprinting" uniquely permits rapid characterization of any RNA or ribonucleoprotein system in solution at single-nucleotide resolution [see, e.g. ref. (*1, 2*)]. Chemical mapping is being advanced by several groups through new approaches for chemical modification, coupling to high-throughput readouts, rapid data processing, high-throughput mutagenesis, and incorporation into structure prediction algorithms (*3–7*).

Perhaps the most widely used RNA chemical probe is dimethyl sulfate (DMS) (*8–11*). DMS modification of the Watson-Crick edge of adenosines or cytosines (at N1 or N3, respectively) blocks reverse transcription, so that reactivities can be obtained by primer extension at single-nucleotide resolution. Nucleotides that appear most strongly protected or reactive to DMS can be inferred to be base-paired or unpaired – qualitative or 'binary' information that can be used for RNA structure modeling by manual or automatic methods (*10, 12*). More recently developed methods, such as selective 2' hydroxyl acylation with primer extension (SHAPE) (*6*), give reactivities that correlate with Watson-Crick base pairing for all nucleotide types, providing more data points than DMS. Indeed, when incorporated into free energy minimization algorithms as pseudo-energy bonuses, SHAPE data can recover RNA secondary structures with high accuracy (*11*); and non-parametric bootstrapping can identify regions with poor confidence (*13*). Nevertheless, this pseudo-energy framework has not been leveraged for prior chemical approaches such as DMS mapping, despite the wide use of these data for in both *in vitro*, *in vivo,* and *in virio* contexts (*9, 12, 14, 15*).

We present herein a benchmark of pseudo-energy-guided secondary structure modeling based on DMS data for 6 non-coding RNAs: unmodified *E. coli* tRNA[phe] (*16*), the P4-P6 domain of the *Tetrahymena* group I ribozyme (*17*), *E. coli* 5S rRNA (*12*), and three ligand-bound domains from bacterial riboswitches [the *V. vulnificus add* adenine riboswitch (*18*), *V. cholerae* cyclic di-GMP riboswitch (*19*), and *F. nucleatum* glycine riboswitch (*20*)]. In all cases, crystallographic data confirmed by solution analyses with the two-dimensional mutate-and-map approach (*21*), have provided 'gold-standard' secondary structures (Supporting Table S1) for evaluating the method's accuracy. The challenging nature of this benchmark was confirmed by the poor accuracy of the *RNAstructure* algorithm without data (Table 1). These models miss 38% of true helices (false negative rate, FNR), and 45% of the returned helices are incorrect (false discovery rate, FDR).

We measured DMS reactivities and estimated errors, inferred from three to eight replicates for each of the six RNAs (Supporting Figures S4 to S9 & Table S1). Analogous to prior SHAPE studies (*11, 13*), we incorporated these DMS data into *RNAstructure* by transforming them into pseudo-energies, giving favorable energies or penalties depending on whether paired nucleotides were DMS-protected or reactive, respectively. We tested pseudo-energy frameworks based on both a previous *ad hoc* formula and an empirically derived statistical potential (inspired by techniques in 3D structure prediction; see Supporting Methods and Figure S1). The two methods gave consistent secondary structures. Because primer extension primarily reads out DMS reactivity at adenosines and cytosines, we excluded reactivities at other bases when performing structure modeling. DMS-guided modeling of the six ncRNAs gave FNR of 9.5% and FDR of 11.6% (Table 1 and Figure 1, see also Table S2), more than three-fold better than without the data. These error rates are lower than those previously achieved by SHAPE-directed modeling [FNR: 17%; FDR: 21% on the same RNAs (*13*)]. Furthermore, the DMS-guided FNR and FDR values are equal to and lower, respectively, than values for SHAPE-based measurements in which primer extension was carried out without deoxyinosine triphosphate (FNR: 9.6%, FDR: 13.6%) to avoid known artefacts (*13*).

We were surprised that DMS mapping gave similar or better information content, compared to SHAPE data, as the latter provides reactivities at approximately twice the number of nucleotides per RNA. Indeed, restricting the algorithm to use SHAPE data at adenines and cytosines or guanines and uracils gave worse models (see Supporting Table S3). Instead, an explanation derives from distinct SHAPE and DMS signatures at nucleotides that are not in Watson-Crick secondary structure but nevertheless form non-canonical interactions (see, e.g., A37 in the *F. nucleatum* glycine riboswitch; Supporting Fig. 2A). These nucleotides appear protected from the SHAPE reaction and thus receive pseudo-energies that incorrectly reward their pairings inside Watson-Crick secondary structure. However, these same nucleotides can expose their Watson-Crick edges to solvent and react strongly with DMS, signifying that they are outside Watson-Crick helices. The DMS-guided modeling can thus return the correct secondary structure in regions where the SHAPE data cannot distinguish Watson-Crick from non-Watson-Crick base pairs (compare Supporting Figs. 2B and 3C).

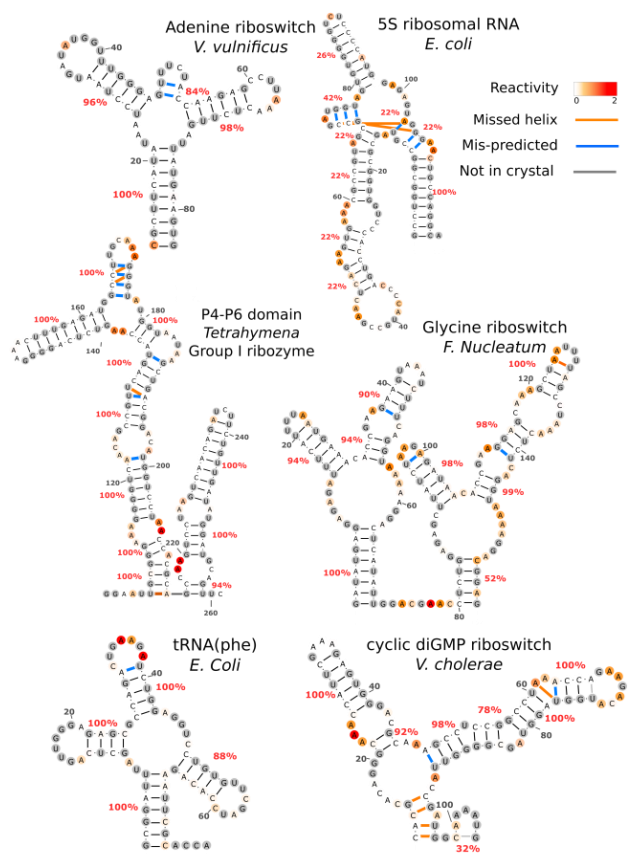

FIGURE 1. Pseudo-energy-guided secondary structure models using DMS data on 6 non-coding RNAs. DMS data and secondary structure models for *E. coli* tRNA$^{phe}$, the P4-P6 domain of the *Tetrahymena* group I ribozyme, *E. coli* 5S rRNA, the *V. vulnificus add* adenine riboswitch, *V. cholerae* cyclic di-GMP riboswitch, and *F. nucleatum* glycine riboswitch. Missed base pairs are highlighted in blue lines; mis-predicted base pairs are indicated by orange lines. Helix bootstrap confidence values are shown in red.

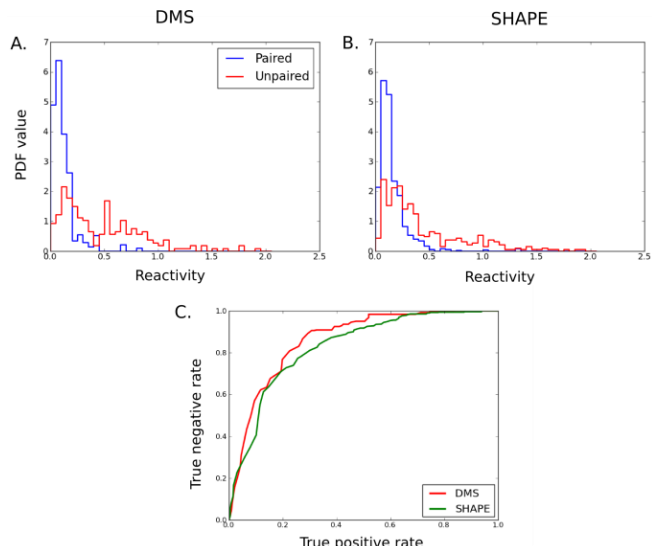

FIGURE 2. Predictive power of DMS and SHAPE. Reactivity histograms for DMS (A) and SHAPE (B). Receiver operating characteristic curves for predicting unpaired nucleotides given a reactivity threshold. Area under the curve (AUC) for DMS is 0.86, for SHAPE, 0.83.

**Table 1: Performance of free energy minimization guided by reactivity-derived pseudo-energies from DMS and SHAPE chemical modifications.**

|  | Total | No data | | DMS | | SHAPE | | DMS + SHAPE | |
|---|---|---|---|---|---|---|---|---|---|
|  |  | TP | FP | TP | FP | TP | FP | TP | FP |
| tRNA$^{phe}$ | 4 | 2 | 3 | 4 | 0 | 4 | 0 | 4 | 0 |
| adenine rbs. | 3 | 2 | 3 | 3 | 1 | 3 | 1 | 3 | 1 |
| cdGMP rbs. | 8 | 6 | 2 | 6 | 0 | 8 | 0 | 8 | 0 |
| 5S rRNA | 7 | 1 | 9 | 6 | 3 | 6 | 3 | 6 | 3 |
| P4-P6 RNA | 11 | 10 | 1 | 10 | 1 | 9 | 1 | 9 | 1 |
| glycine rbs. | 9 | 5 | 3 | 9 | 0 | 8 | 0 | 9 | 0 |
| Total | 42 | 26 | 21 | 38 | 6 | 38 | 5 | 39 | 5 |
| FNR | | 38.1% | | 9.5% | | 9.5% | | 7.1% | |
| FDR | | 44.7% | | 11.6% | | 13.6% | | 11.4% | |
| Sensitivity | | 61.9% | | 90.5% | | 90.5% | | 92.9% | |
| PPV | | 55.3% | | 88.4% | | 86.4% | | 88.6% | |

Abreviations: TP, true positives; FP false positives; Cryst., number of helices in crystallographic model; FNR, False negative rate = 1 – TP/Cryst.; FDR, False discovery rate = FP/(TP + FP); Sensitivity = (1 – FNR); PPV, Positive predictive value = (1 – FDR)

Reactivity histograms (Figure 2A and 2B) further support the enhanced predictive power of DMS vis-à-vis SHAPE. DMS mapping better distinguishes between nucleotides inside Watson-Crick helices and nucleotides outside helices (see also receiver operating characteristic curve; Figure 2C.)

Like SHAPE-guided modeling, DMS-directed structure inference still produces errors (Table 1), e.g., for the central junction of the 5S rRNA (Supporting Fig. 2E and 2F). Some of these errors may be resolved through better incorporation of the DMS-derived pseudoenergies at, e.g., 'singlet' base pairs (Supporting Fig. 2E). Nevertheless, as with SHAPE modeling, these erroneous regions can be pinpointed by estimating helix-by-helix confidence values through non-parametric boostrapping [Supplemental Methods and ref. (*13*); see also Supporting Figure S3]. For example, this procedure gives high confidence (≥ 90%) at almost all helices in the glycine riboswitch but low confidence values (<50 %) throughout the 5S rRNA DMS model (Figure 1).

For many applications, DMS and SHAPE measurements can be carried out in parallel, so we sought to determine if their combination might improve automated secondary structure inference. Application of both sets of pseudo-energies gave a slight improvement in the algorithm's accuracy (FNR of 7.1% and FDR of 11.4% ). In addition, we performed measurements with a reagent that primarily modifies Waston/Crick edges of guanosine and uracil, 1-cyclohexyl-(2-morpholinoethyl) carbodiimide metho-p-toluene sulfonate (CMCT) (*22*). Incorporation of these data into *RNAstructure* gave poorer accuracy modeling than the DMS- or SHAPE-guided modeling above (FNR of 14.3%, FDR or 18.2%; see Supporting Table S4), consistent with weaker discrimination between paired/unpaired residues (Supporting Figure S1); and integrating CMCT with DMS and/or SHAPE data did not improve accuracy (Supplemental Table S2).

The benchmark results presented herein establish that chemical mapping with dimethyl sulfate (DMS) can achieve prediction accuracies comparable to the SHAPE protocol using pseudo-energies to guide free energy minimization. DMS has been extensively used both *in vitro* and *in vivo*, for time-resolved RNA folding, precise thermodynamic analysis, and mapping RNA/protein interfaces (*9*, *12*, *14*, *15*, *22*). Sophisticated techniques for optimizing the reaction rate and its quenching have been developed (*9*, *23*). Applying automated structure modeling, as demonstrated herein, will enable researchers to better take advantage of this large body of previous work. Furthermore, future studies may find it advantageous to perform both DMS and SHAPE approaches in parallel. Along with bootstrapping (*13*), comparison of separate DMS-guided vs. SHAPE-guided secondary structure models will permit rapid assessment of systematic errors and thus provide more accurate inferences.

## ACKNOWLEDGEMENT

We thank authors of *RNAstructure* for making the source code freely available and members of the Das lab for comments on the manuscript.## SUPPORTING INFORMATION

Supporting methods, figures, and model accuracy tables are available free of charge at http://pubs.acs.org.

## REFERENCES

1. Black, D. L., and Pinto, A. L. (1989) U5 small nuclear ribonucleoprotein: RNA structure analysis and ATP-dependent interaction with U4/U6., *Molecular and cellular biology 9*, 3350-9.
2. Moazed, D., and Noller, H. F. (1991) Sites of interaction of the CCA end of peptidyl-tRNA with 23S rRNA., *Proceedings of the National Academy of Sciences of the United States of America 88*, 3725-8.
3. Mitra, S., Shcherbakova, I. V., Altman, R. B., Brenowitz, M., and Laederach, A. (2008) High-throughput single-nucleotide structural mapping by capillary automated footprinting analysis., *Nucleic acids research*. Oxford University Press *36*, e63.
4. Yoon, S., Kim, J., Hum, J., Kim, H., Park, S., Kladwang, W., and Das, R. (2011) HiTRACE: high-throughput robust analysis for capillary electrophoresis, *Bioinformatics 27*, 1798-1805.
5. Kladwang, W., Cordero, P., and Das, R. (2011) A mutate-and-map strategy accurately infers the base pairs of a 35-nucleotide model RNA, *RNA 17*, 522-534.
6. Wilkinson, K. A., Merino, E. J., and Weeks, K. M. (2006) Selective 2'-hydroxyl acylation analyzed by primer extension (SHAPE): quantitative RNA structure analysis at single nucleotide resolution., *Nature protocols*. Nature Publishing Group *1*, 1610-6.
7. Lucks, J. B., Mortimer, S. A., Trapnell, C., Luo, S., Aviran, S., Schroth, G. P., Pachter, L., Doudna, J. A., and Arkin, A. P. (2011) Multiplexed RNA structure characterization with selective 2'-hydroxyl acylation analyzed by primer extension sequencing (SHAPE-Seq)., *Proceedings of the National Academy of Sciences of the United States of America* (Hage, J., and Meeus, M., Eds.). National Academy of Sciences *108*, 11063-11068.
8. Peattie, D. A., and Gilbert, W. (1980) Chemical probes for higher-order structure in RNA., *Proceedings of the National Academy of Sciences of the United States of America*. National Acad Sciences *77*, 4679-4682.
9. Tijerina, P., Mohr, S., and Russell, R. (2007) DMS footprinting of structured RNAs and RNA-protein complexes., *Nature Protocols 2*, 2608-2623.
10. Mathews, D. H., Disney, M. D., Childs, J. L., Schroeder, S. J., Zuker, M., and Turner, D. H. (2004) Incorporating chemical modification constraints into a dynamic programming algorithm for prediction of RNA secondary structure., *Proceedings of the National Academy of Sciences of the United States of America*. National Academy of Sciences *101*, 7287-92.
11. Deigan, K. E., Li, T. W., Mathews, D. H., and Weeks, K. M. (2009) Accurate SHAPE-directed RNA structure determination., *Proceedings of the National Academy of Sciences of the United States of America*. National Academy of Sciences *106*, 97-102.
12. Leontis, N. B., and Westhof, E. (1998) The 5S rRNA loop E: Chemical probing and phylogenetic data versus crystal structure, *RNA 4*, 1134-1153.
13. Kladwang, W., VanLang, C. C., Cordero, P., and Das, R. (2011) Understanding the errors of SHAPE-directed RNA structure modeling., *Biochemistry*. American Chemical Society *50*, 8049-56.
14. Lempereur, L., Nicoloso, M., Riehl, N., Ehresmann, C., Ehresmann, B., and Bachellerie, J. P. (1985) Conformation of yeast 18S rRNA. Direct chemical probing of the 5′ domain in ribosomal subunits and in deproteinized RNA by reverse transcriptase mapping of dimethyl sulfate-accessible sites, *Nucleic Acids Research*. Oxford Univ Press *13*, 8339.
15. Wells, S. E., Hughes, J. M., Igel, A. H., and Ares, M. (2000) Use of dimethyl sulfate to probe RNA structure in vivo., *Methods in Enzymology*. Academic Press *318*, 479-493.
16. Byrne, R. T., Konevega, A. L., Rodnina, M. V., and Antson, A. A. (2010) The crystal structure of unmodified tRNAPhe from Escherichia coli, *Nucleic Acids Research*. Oxford University Press *38*, 4154-4162.
17. Cate, J. H., Gooding, A. R., Podell, E., Zhou, K., Golden, B. L., Kundrot, C. E., Cech, T. R., and Doudna, J. A. (1996) Crystal Structure of a Group I Ribozyme Domain: Principles of RNA Packing, *Science 273*, 1678-1685.
18. Serganov, A., Yuan, Y.-R., Pikovskaya, O., Polonskaia, A., Malinina, L., Phan, A. T., Hobartner, C., Micura, R., Breaker, R. R., and Patel, D. J. (2004) Structural basis for discriminative regulation of gene expression by adenine- and guanine-sensing mRNAs., *Chemistry & biology 11*, 1729-41.
19. Smith, K. D., Lipchock, S. V., Livingston, A. L., Shanahan, C. A., and Strobel, S. A. (2010) Structural and Biochemical Determinants of Ligand Binding by the c-di-GMP Riboswitch,, *Biochemistry*. American Chemical Society *49*, 7351-7359.
20. Butler, E. B., Xiong, Y., Wang, J., and Strobel, S. A. (2011) Structural basis of cooperative ligand binding by the glycine riboswitch., *Chemistry & biology 18*, 293-8.
21. Kladwang, W., VanLang, C. C., Cordero, P., and Das, R. (2011) A two-dimensional mutate-and-map strategy for non-coding RNA structure., *Nature chemistry*. Nature Publishing Group *3*, 954-62.
22. Planning, S. (2000) Probing RNA Structure with Chemical, *Current protocols in nucleic acid chemistry edited by Serge L Beaucage et al Chapter 6*, 1-21.

23. Das, R., Karanicolas, J., and Baker, D. (2010) Atomic accuracy in predicting and designing noncanonical RNA structure., *Nature methods*. Nature Publishing Group 7, 291-4.

Table of Contents artwork

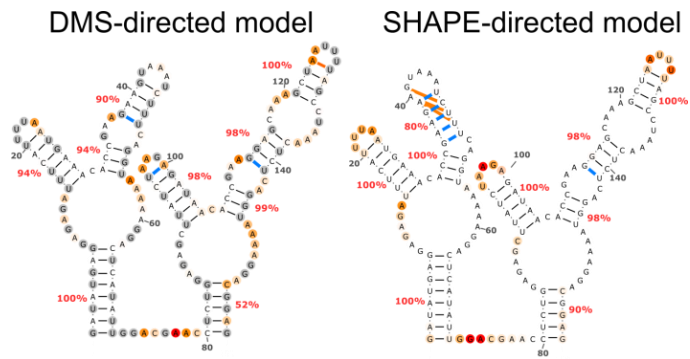

# Supporting information

*for the manuscript*

**Quantitative DMS mapping for automated RNA secondary structure inference**

Pablo Cordero[1], Wipapat Kladwang[2], Christopher VanLang[3], and Rhiju Das[1,2,4*]
Departments of Biomedical Informatics[1], Biochemistry[2], Chemical Engineering[3], and Physics[4],
Stanford University, Stanford CA 94305

**Tables and Figures contained in this document:**

Supporting Methods
Table S1: RNA systems used in this study
Table S2: Base-pair-wise accuracy table for DMS and SHAPE
Table S3: Using only SHAPE reactivities in adenines and cytosines does not improve structure modeling
Table S4: Helix-wise accuracy table for inclusion of CMCT in structure modeling.
Figure S1: A probabilistic potential for pseudoenergy bonuses
Figure S2. DMS vis-à-vis SHAPE for secondary structure inference
Figure S3: Histogram of helix bootstrap values for SHAPE, DMS, and CMCT
Figure S4. DMS, SHAPE, and CMCT data and pseudo-energy guided models for the *add* adenine riboswitch
Figure S5. DMS, SHAPE, and CMCT data and pseudo-energy guided models for tRNA$^{phe}$
Figure S6. DMS, SHAPE, and CMCT data and pseudo-energy guided models for the cyclic di-GMP riboswitch
Figure S7. DMS, SHAPE, and CMCT data and pseudo-energy guided models for the 5S rRNA
Figure S8. DMS, SHAPE, and CMCT data and pseudo-energy guided models for the P4-P6 domain of the *Tetrahymena* group I ribozyme
Figure S9. DMS, SHAPE, and CMCT data and pseudo-energy guided models for the *F. nucleatum* glycine riboswitch

## Supporting Methods

**Experimental setup**

Chemical mapping experiments were performed using *in vitro* transcribed RNAs from PCR-assembled DNA templates as previously described (*1*). All SHAPE, DMS, and CMCT measurements were performed at least in triplicate using three independent RNA preparations. DNA templates containing a T7 RNA polymerase promoter sequence (TTCTAATACGACTCACTATA) followed by the sequence of interest and a reverse transcription primer binding site (AAACAAACAACAACAACAAC) were PCR-assembled from oligomers of up to 60 nucleotides in length (Integrated DNA Technologies) with a Phusion DNA polymerase (Finnzymes) and purified with AMPure magnetic beads (Agencourt, Beckman Coulter). Sample concentrations were calculated through UV absorbances on a Nanodrop 100 spectrophotometer and lengths verified in 4% agarose gels. *In vitro* RNA transcription was performed as previously described using a T7 RNA polymerase (New England Biolabs) and purified with MagMax magnetic beads (Ambion) or an RNA clean kit (Zymo research); RNA from the two purification methods gave indistinguishable results. RNA concentrations were measured on a Nanodrop 100 spectrophotometer.

Chemical modification was performed in volumes of 20 μL with 1.2 pmols of RNA in 50 mM Na-HEPES (pH 8.0), 10 mM $MgCl_2$, ligand at the desired concentration for riboswitches (see Table S1) and 5 μL of modification reagent [1% (10.5 mM) dimethyl sulfate (DMS) prepared by mixing 10 uL 10.5 M DMS into 90 μL ethanol, and then 900 μL water; 42 mg/mL (99 mM) 1-cyclohexyl-(2-morpholinoethyl) carbodiimide metho-p-toluene sulfonate (CMCT); or 24 mg/mL (0.14 mM) N-methylisatoic anhydride (NMIA)]. Modification reactions were incubated at 24 °C for 15 to 60 minutes depending on the length of the RNA to achieve overall modification rates of less than 30% and then quenched appropriately (adding 5 μl of 0.5 M Na-MES, pH 6.0 for SHAPE and CMCT or 2-mercaptoethanol for DMS). Quenches also included 1μL of poly(dT) magnetic beads (Ambion) and

0.065 pmols of 5′-rhodamine-green-labeled primer
(AAAAAAAAAAAAAAAAAAAAGTTGTTGTTGTTGTTTCTTT) complementary to the 3′ end of the
RNAs used for reverse transcription. The reaction mixtures were purified by magnetic separation,
rinsed with 40 µL of 70% ethanol twice, and allowed to air-dry for 10 min while sitting in the magnetic
post stand. The magnetic bead mixtures were resuspended in 2.5 µL of deionized water and reverse
transcribed by adding a premix solution containing 0.2 µL of SuperScript III (Invitrogen), 1.0 µL of 5×
SuperScript First Strand buffer (Invitrogen), 0.4 µL of dNTPs at 10 mM each (dATP, dCTP, dGTP, and
dTTP; dITP was not used to reverse-transcribe these RNAs, as it generates signal artifacts in NMIA
chemical mapping protocols (*1*)), 0.25 µL of 0.1 M dithiothreitol (DTT), and 0.65 µL of water and
incubating at 42 °C for 30 min. RNA was hydrolyzed by adding 5 µL of 0.4 M NaOH and incubating at
90 °C for 3 min. The solutions were neutralized by the addition of 5 µL of an acid quench (2 volumes
of 5 M NaCl, 2 volumes of 2 M HCl, and 3 volumes of 3 M sodium acetate) and the resulting
fluorescent DNA was purified by magnetic bead separation. The beads were washed with 40 µL of
70% ethanol, air-dried for 5 minutes, and resuspended in 10 µL of a solution containing 0.125 mM Na-
EDTA (pH 8.0) and a Texas Red-labeled reference ladder, or in 10 µL of HiDi formamide containing a
ROX 350 ladder (Applied Biosystems) spectrally distinct from the rhodamine-green chemical mapping
signal. For verifying sequence assignments, reference ladders were created using an analogous
protocol without chemical modification and including 2′,3′-dideoxy-GTP in an amount equimolar with
dGTP during reverse transcription. The products were separated by capillary electrophoresis on an
ABI 3100 or ABI 3700 DNA sequencer.

**Analysis of electropherogram traces and quantification of reactivities**

Electropherograms were analyzed with the HiTRACE software (*2*). Sequence assignments for bands
were obtained by alignment to sequencing lanes with incorporated ddATP, ddCTP, ddGTP, or ddTTP
nucleotides. Band intensities were fit as Gaussian peaks and processed through a likelihood-based
framework for overmodification correction and background subtraction as defined previously through
the *overmod_and_background_correct_logL.m* and *get_average_standard_state.m* HiTRACE scripts

(*1*). For DMS and CMCT data, reactivities at guanines/uracils and adenines/cytosines, respectively, are expected to be low. Therefore, when performing likelihood-based background substraction on data for those nucleotides, we used a distribution of the form $p(x) = \exp(F(x))$ where $F(x) = F_+(x - x_0)$ if $x > x_0$ and $F(x) = -F_-(x - x_0)$ otherwise, with parameter values $F_- = F_+ = 25$, $x_0 = 0$ (for details on this functional form see ref (*1*)). This distribution corresponds to positions with lower expected reactivities, thus attenuating the final reactivity values for those nucleotides.

Final averaged data and errors have been made publicly available in the RNA Mapping Database (http://rmdb.stanford.edu). Accession IDs corresponding to each modifier are:

- For NMIA (SHAPE) modified samples, entries TRNAPH_SHP_0003, TRP4P6_SHP_0004, 5SRRNA_SHP_0003, ADDRSW_SHP_0004, CIDGMP_SHP_0003, and GLYCFN_SHP_0006 were submitted.
- For DMS modified samples, new entries TRNAPH_DMS_0001, TRP4P6_DMS_0001, 5SRRNA_DMS_0001, ADDRSW_DMS_0001, CIDGMP_DMS_0001, and GLYCFN_DMS_0001 were submitted.
- For CMCT modified samples, new entries TRNAPH_CMC_0001, TRP4P6_CMC_0001, 5SRRNA_CMC_0001, ADDRSW_CMC_0001, CIDGMP_CMC_0001, and GLYCFN_CMC_0001 were submitted.

**Computational methods**

We tested the modeling accuracy of minimum free energy structure calculation with reactivity-derived pseudo-energies added to the scoring function (*3*). The *Fold* executable of the RNAstructure package (version 5.3) was used to infer pseudo-energy-directed secondary structure models. Pseudo-energies were applied once for each nucleotide that forms an edge base pair and twice for each nucleotide that forms an internal base pair. Additionally, non-parametric bootstrap analysis was performed to estimate helix-wise prediction confidence (*1*).

Previous work used an energy-like functional form with two free parameters to calculate pseudo-energies from experimental chemical mapping reactivities that are given as bonuses or penalties to the energy scoring function of a secondary structure prediction algorithm ($\Delta G_i = m \log(S_i + 1) + b$, $S_i$ is the reactivity value at position *i*; *m* and *b* are free parameters, see ref (*3*)). We also tested a more direct way of expressing the pseudo-energy potential by taking the log-likelihood ratio of a base being unpaired versus paired given a chemical reactivity value:

$$\Delta G = -k_B T \log\left(\frac{P(S_i \mid i \text{ is paired})}{P(S_i \mid i \text{ is unpaired})}\right)$$

Here, *T* is the temperature and $k_B$ is the Boltzmann constant. The likelihoods for paired and unpaired reactivities were derived from a mixture of two gamma distributions to reactivities of paired and unpaired nucleotides in our non-coding RNA benchmark (see Figure S4). This probabilistic potential is akin to those found in forcefields that include knowledge-based terms, such as the ROSETTA framework for three-dimensional structure modeling (*4–6*). In the future, if different reactivity distributions are discovered for different features (e.g., apical loops and interior loops), this framework permits the facile incorporation of that information.

We applied our probabilistic potential to calculate pseudo-energies to guide the free-energy minimization *Fold* program in the RNAstructure package. The performance of the algorithm using this probabilistic potential for SHAPE, DMS, and CMCT reactivities is given in Table S6 and is identical to results obtained for the standard potential with slope (*m*) and intercept (*b*) optimized through grid-search. To test for over-fitting, we performed leave-one-out-validations for each RNA by fitting the probabilistic potential using the mapping data of the other RNAs and re-running the algorithm; validation results were identical to those when using the full data. RNAstructure was modified to allow the DMS and CMCT data to be input through the flags *-dms* and *-cmct*. We are in the process of contacting the authors of RNAstructure to include these options in the next release.

**Assessment of accuracy**

We evaluated the predictions as defined previously in refs (*1*, *7*): a crystallographic helix was considered correctly recovered if more than 50% of its base pairs were observed in a helix by the computational model; ±1 helix shifts were not considered correct. Modeling errors are expressed as false negative rates (FNR; fraction of crystallographic helices that were predicted to be single-stranded) and false discovery rates (FDR; fraction of predicted helices that were not present in crystallographic models). We also include positive predictive values (PPV) and sensitivities of each approach, and all metrics at the level of individual base pairs rather than helices (see Table S6).

**Table S1: RNA systems used in this study**

| RNA, source | Solution conditions[a] | Replicates | Experiments | Offset[b] | PDB[c] |
|---|---|---|---|---|---|
| tRNAphe, *E. coli* | Standard | SHAPE: 6<br>DMS: 5<br>CMCT: 4 | SHAPE: 4<br>DMS: 4<br>CMCT: 3 | -15 | **1TRA**<br>1EHZ |
| P4-P6 domain, *Tetrahymena* ribozyme | Standard | SHAPE: 11<br>DMS: 11<br>CMCT: 4 | SHAPE: 5<br>DMS: 5<br>CMCT: 3 | 89 | **1GID**<br>1L8V<br>1HR2<br>2R8S |
| 5S rRNA, *E. coli* | Standard | SHAPE: 5<br>DMS: 5<br>CMCT: 3 | SHAPE: 3<br>DMS: 4<br>CMCT: 3 | -20 | **3OFC**<br>3OAS<br>3ORB<br>2WWQ<br>… |
| Adenine riboswitch, *V. vulnificus* (*add*) | Standard + 5 mM adenine | SHAPE: 4<br>DMS: 3<br>CMCT: 3 | SHAPE: 3<br>DMS: 3<br>CMCT: 3 | -8 | **1Y26**<br>1Y27<br>2G9C<br>3GO2<br>… |
| c-di-GMP riboswitch, *V. cholerae* (*VC1722*) | Standard + 10 µM cyclic di-guanosine monophosphate | SHAPE: 5<br>DMS: 3<br>CMCT: 3 | SHAPE: 3<br>DMS: 2<br>CMCT: 3 | 0 | **3MXH**<br>3IWN<br>3MUV<br>3MUT<br>… |
| Glycine riboswitch, *F. nucleatum* | Standard + 10 mM glycine | SHAPE: 16<br>DMS: 7<br>CMCT: 9 | SHAPE: 4<br>DMS: 3<br>CMCT: 3 | -10 | **3P49** |

[a] Standard conditions were: 10 mM $MgCl_2$, 50 mM Na-HEPES, pH 8.0 at 24 °C.
[b] Offset added to the original numbering scheme of the sequence from which this subsequence was taken from.
[c] Boldfaced IDs correspond to the PDB entries from which the sequence for this study was taken. Additional PDB entries correspond to other studies of the same RNA system.

| RNA, source | Sequence[d] | Secondary Structure[e] |
|---|---|---|
| tRNAphe, *E. coli* | ggaacaaacaaaacaGCGGAUUUAGCUCAGUUGGGAGAGCGCCAGACUGAAGAUCUGGAGGUCCUGUGUUCGAUCCACAGAAUUCGCACCAaaaccaaagaaacaacaacaacaac | .............((((((((..(((((........)))).((((.........))))....(((((((.......))))))))))))))))................ |
| P4-P6 domain, *Tetrahymena* ribozyme | ggccaaaacaacgGAAUUGCGGGAAAGGGGUCAACAGCCGUUCAGUACCAAGUCUCAGGGGAAACUUUGAGAUGGCCUUGCAAAGGGUAUGGUAAUAAGCUGACGGACAUGGUCCUAACCACGCAGCCAAGUCCUAAGUCAACAGAUCUUCUGUUGAUAUGGAUGCAGUUCAaaaccaaaccaaagaaacaacaacaacaac | ...............(((((...(((((......(((.((((.(((..(((((((((....))))))))))..((........))..))).....)))))))..)))))))..)).))))((...((((...((((((((...))))))))..))))...))........................ |
| 5S rRNA, *E. coli* | ggaaaggaaagggaaagaaaUGCCUGGCGGCCGUAGCGCGGUGGUCCCACCUGACCCCAUGCCGAACUCAGAAGUGAAACGCCGUAGCGCCGAUGGUAGUGUGGGGUCUCCCCAUGCGAGAGUAGGGAACUGCCAGGCAUaaaacaaaacaaagaaacaacaacaacaac | ..............((((((((......(((((((((....((((((((.............))))..)))...))))))).)).(((.......(((((((((...))))))))).......))...))))))))))............................ |
| Adenine riboswitch, *V. vulnificus* (*add*) | ggaaaggaaagggaaagaaaCGCUUCAUAUAAUCCUAAUGAUAUGGUUUGGGAGUUUCUACCAAGAGCCUUAAACUCUUGAUUAUGAAGUGaaaacaaaacaaagaaacaacaacaacaac | ....................((((((((...(((((.........)))))........(((((........))))))..))))))))............................. |
| c-di-GMP riboswitch, *V. cholerae* (*VC1722*) | ggaaaaauGUCACGCACAGGGCAAACCAUUCGAAAGAGUGGGACGCAAAGCCUCCGGCCUAAACCAGAAGACAUGGUAGGUAGCGGGGUUACCGAUGGCAAAAUGcauacaaaccaaagaaacaacaacaacaac | ..........((((......((...((((((....))))))...))...(((.((((((((..((..........))))))))..))))...)).)) ..................... |
| Glycine riboswitch, *F. nucleatum* | ggacagagagGAUAUGAGGAGAGAUUUCAUUUUAAUGAAACACCGAAGAAGUAAAUCUUUCAGGUAAAAAGGACUCAUAUUGGACGAACCUCUGGAGAGCUUAUCUAAGAGAUAACACCGAAGGAGCAAAGCUAAUUUUAGCCUAAACUCUCAGGUAAAAGGACGGAgaaaacacaacaaagaaacaacaacaacaac | ..........(((((((((......(((((....)))))).(((....(((.....)))...)))........))))))))).........(((((....((((((.....)))))).(((...((((......(((....)))......)))))..)))......)))))............................. |

[d] Lowercase subsequences are buffer sequences and primer binding sites used for transcription and reverse transcription. These additional sequences do not interfere with the RNAs' structures according to ViennaRNA and RNAstructure predictions.

[e] Structure is given dot-bracket notation. A two-base pair helix for the adenine riboswitch that is not nested is not included in this dot-bracket representation.

**Table S2: Base-pair-wise accuracy table for SHAPE and DMS**

|  | Total | no data | | DMS | | SHAPE | | DMS + SHAPE | |
|---|---|---|---|---|---|---|---|---|---|
|  |  | TP | FP | TP | FP | TP | FP | TP | FP |
| tRNA[phe] | 20 | 12 | 12 | 20 | 1 | 20 | 1 | 20 | 1 |
| adenine rbsw. | 21 | 15 | 10 | 21 | 2 | 21 | 2 | 21 | 2 |
| cyclic di-GMP rbsw. | 25 | 21 | 5 | 21 | 1 | 25 | 2 | 25 | 1 |
| 5S rRNA | 34 | 9 | 31 | 32 | 6 | 32 | 6 | 32 | 6 |
| P4-P6 RNA | 48 | 44 | 9 | 45 | 6 | 44 | 6 | 44 | 5 |
| glycine rbsw. | 40 | 23 | 18 | 40 | 2 | 37 | 6 | 40 | 2 |
| **Total** | 188 | 124 | 85 | 179 | 18 | 179 | 23 | 182 | 17 |
| **FNR** |  | 34% | | 4.8% | | 4.8% | | 3.2% | |
| **FDR** |  | 40.7% | | 9.1% | | 11.4% | | 8.5% | |
| **Sensitivity** |  | 66% | | 95.2% | | 95.2% | | 96.8% | |
| **PPV** |  | 59.3% | | 91% | | 88.6% | | 91.5% | |

**Table S3: Using only SHAPE reactivities in adenines and cytosines does not improve structure modeling** – To test if the quality of the models given by DMS could be explained by selectively applying pseudo-energies only to adenines and cytosines, we re-ran the *Fold* program only with SHAPE reactivities that fell in adenines and cytosines. The resulting models have worse FDR and FNR than those derived from using DMS or full SHAPE data, confirming that the DMS results could not be explained by applying pseudo-energies to a subset of positions in the RNA. The reported accuracies are helix-wise.

|  | Total | no data | | SHAPE As and Us | | SHAPE Gs and Us | | SHAPE | |
|---|---|---|---|---|---|---|---|---|---|
|  |  | TP | FP | TP | FP | TP | FP | TP | FP |
| tRNA[phe] | 4 | 2 | 3 | 4 | 0 | 4 | 0 | 4 | 0 |
| adenine rbsw. | 3 | 2 | 3 | 3 | 0 | 3 | 0 | 3 | 1 |
| cyclic di-GMP rbsw. | 8 | 6 | 2 | 5 | 2 | 6 | 1 | 8 | 0 |
| 5S rRNA | 7 | 1 | 9 | 1 | 7 | 2 | 5 | 6 | 3 |
| P4-P6 RNA | 11 | 10 | 1 | 9 | 1 | 8 | 2 | 9 | 1 |
| glycine rbsw. | 9 | 5 | 3 | 8 | 1 | 8 | 1 | 8 | 1 |
| **Total** | 42 | 26 | 21 | 30 | 11 | 31 | 9 | 38 | 6 |
| **FNR** |  | 38.1% | | 28.6% | | 26.2% | | 9.5% | |
| **FDR** |  | 44.7% | | 26.8% | | 22.5% | | 13.6% | |
| **Sensitivity** |  | 61.9% | | 71% | | 73.8% | | 90.5% | |
| **PPV** |  | 55.3% | | 73.2% | | 77.5% | | 86.4% | |

**Table S4: Helix-wise accuracies for inclusion of CMCT data in structure modeling.**

|  | Total | no data | | CMCT | | CMCT + DMS | | CMCT + SHAPE | | CMCT + DMS + SHAPE | |
|---|---|---|---|---|---|---|---|---|---|---|---|
|  |  | TP | FP | TP | FP | TP | FP | TP | FP | TP | FP |
| tRNA[phe] | 4 | 2 | 3 | 4 | 0 | 4 | 0 | 4 | 0 | 4 | 0 |
| adenine rbsw. | 3 | 2 | 3 | 3 | 1 | 3 | 1 | 3 | 1 | 3 | 1 |
| cyclic di-GMP rbsw. | 8 | 6 | 2 | 5 | 2 | 6 | 0 | 6 | 2 | 8 | 0 |
| 5S rRNA | 7 | 1 | 9 | 6 | 3 | 6 | 3 | 6 | 3 | 6 | 3 |
| P4-P6 RNA | 11 | 10 | 1 | 10 | 1 | 10 | 1 | 9 | 1 | 9 | 1 |
| glycine rbsw. | 9 | 5 | 3 | 8 | 1 | 9 | 0 | 8 | 1 | 9 | 0 |
| **Total** | 42 | 26 | 21 | 36 | 8 | 38 | 5 | 36 | 8 | 39 | 5 |
| **FNR** |  | 38.1% | | 14.3% | | 9.5% | | 14.3% | | 7.1% | |
| **FDR** |  | 44.7% | | 18.2% | | 11.6% | | 18.2% | | 11.4% | |
| **Sensitivity** |  | 61.9% | | 85.7% | | 90.5% | | 85.7% | | 92.9% | |
| **PPV** |  | 55.3% | | 81.8% | | 88.4% | | 81.8% | | 88.6% | |

**Figure S1. A probabilistic potential for pseudoenergy bonuses –** (A) Normalized histograms for paired and unpaired reactivities (as defined by the crystallographic model) are retrieved from chemical mapping data. (B) Gamma mixture distributions with two components fitted to the data (dashed lines) and the pseudo-energies are then calculated as a function of the log-likelihood ratio of paired and unpaired distributions [$\Delta G_i = -k_\beta T \log(P(S_i \mid i$ is paired$)/P(S_i \mid i$ is unpaired$))$ for every $S_i$ reactivity at nucleotide $i$]; error estimates are calculated by a smooth bootstrap procedure. Optimizing the slope ($m$) and intercept ($b$) of the standard potential ([$\Delta G_i = m \log(S_i +1) + b$) gives identical results than the aforementioned probabilistic approach.

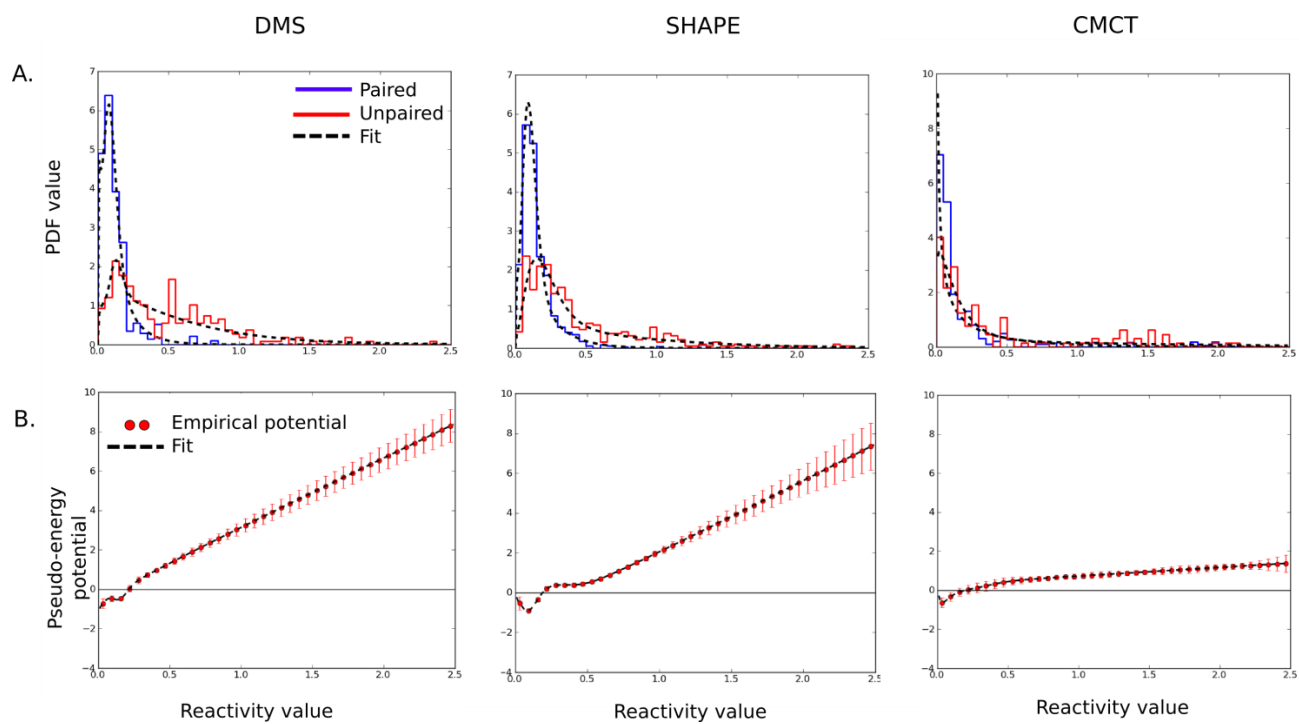

**Figure S2. DMS vis-à-vis SHAPE for secondary structure inference** – (A) A37 (green) in the glycine riboswitch has its Watson-Crick edge exposed, making its N1 atom (red sphere) accessible to DMS modification, guiding *RNAstructure* to the correct helix (B). However, A37 is stabilized by local interactions, protecting it from SHAPE chemistries; the algorithm thus is mislead into an incorrect helix (C). Pseudo-energy-guided models fail to correctly infer the structure of the 5S rRNA (D); both DMS (E) and SHAPE (F) guide the secondary structure calculation to an incorrect central junction. Only two Watson-Crick base pairs support the correct helix (orange, D-F). Further, current implementation of the pseudo-energy framework does not apply the bonuses/penalties to 'singlet' base-pairs, allowing the formation of incorrect, short helices (see e.g. E and F). Secondary structure figures were prepared in VARNA (*8*).

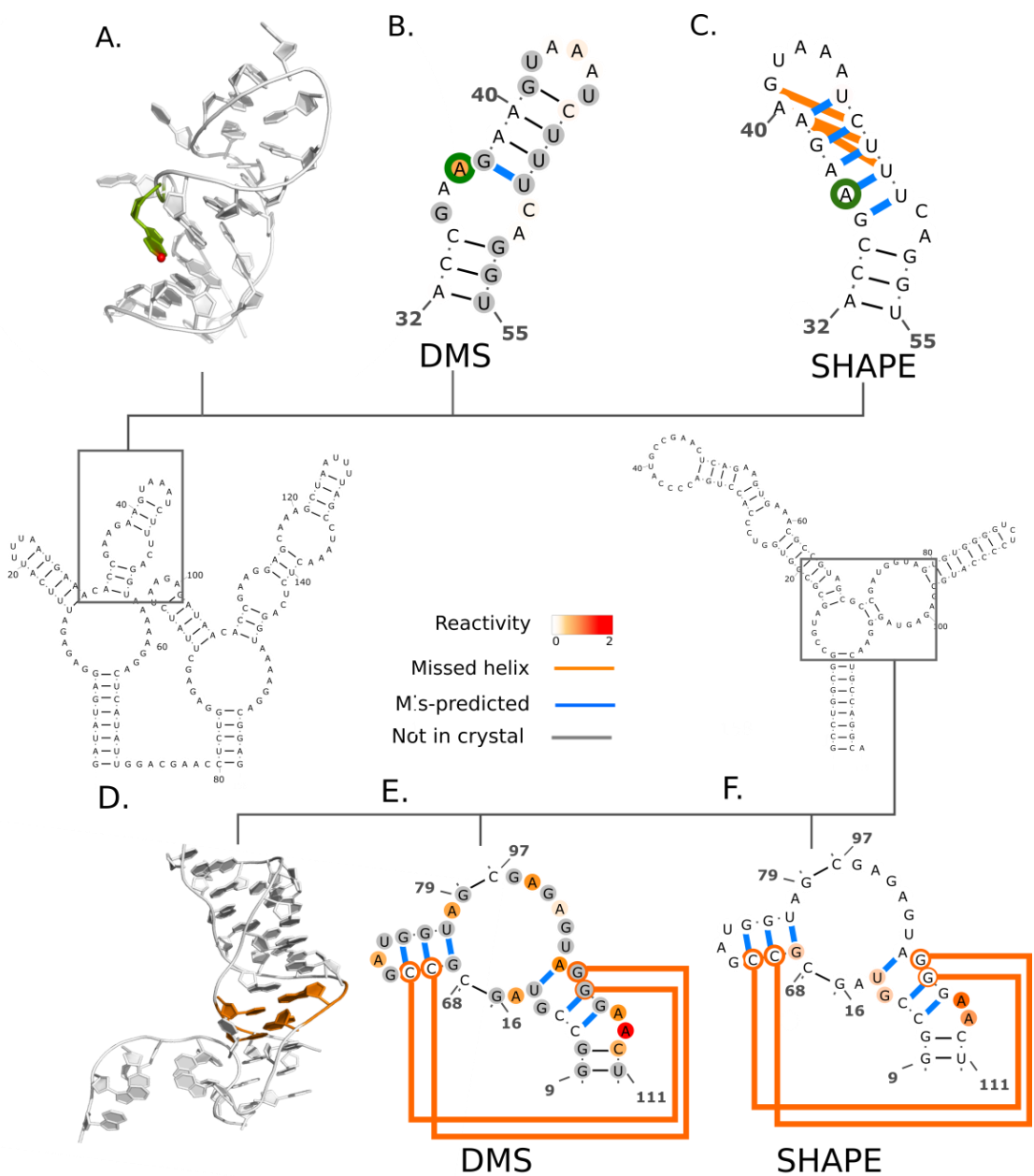

**Figure S3. Helix-wise, bootsrap confidence value histograms for DMS, SHAPE, and CMCT models. –** Blue histograms are for correctly predicted helices (true positives), red histograms are for incorrectly predicted helices (false positives).

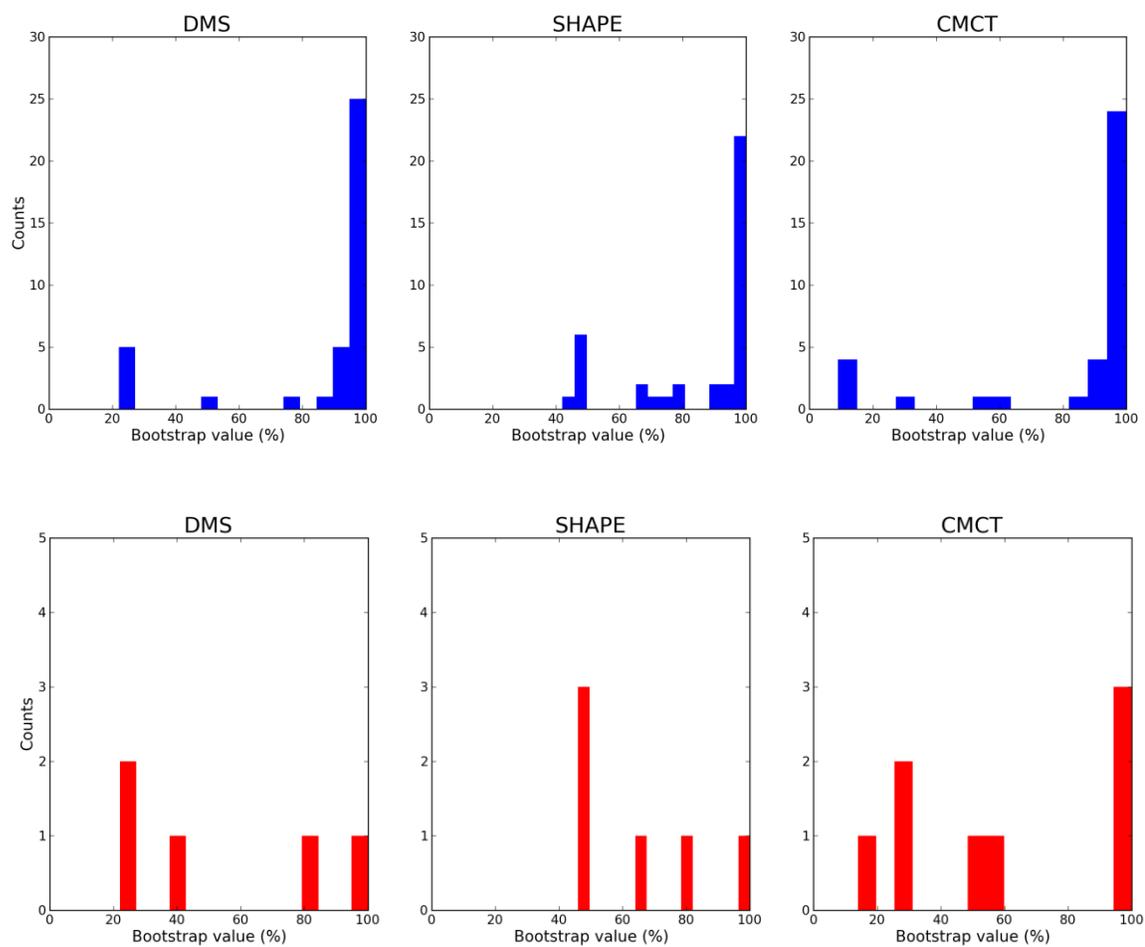

**Figure S4. DMS, SHAPE, and CMCT data and pseudo-energy guided models for the *add* adenine riboswitch –** DMS and CMCT data at guanines and uracils, and adenines and cytosines, respectively, are marked in gray.

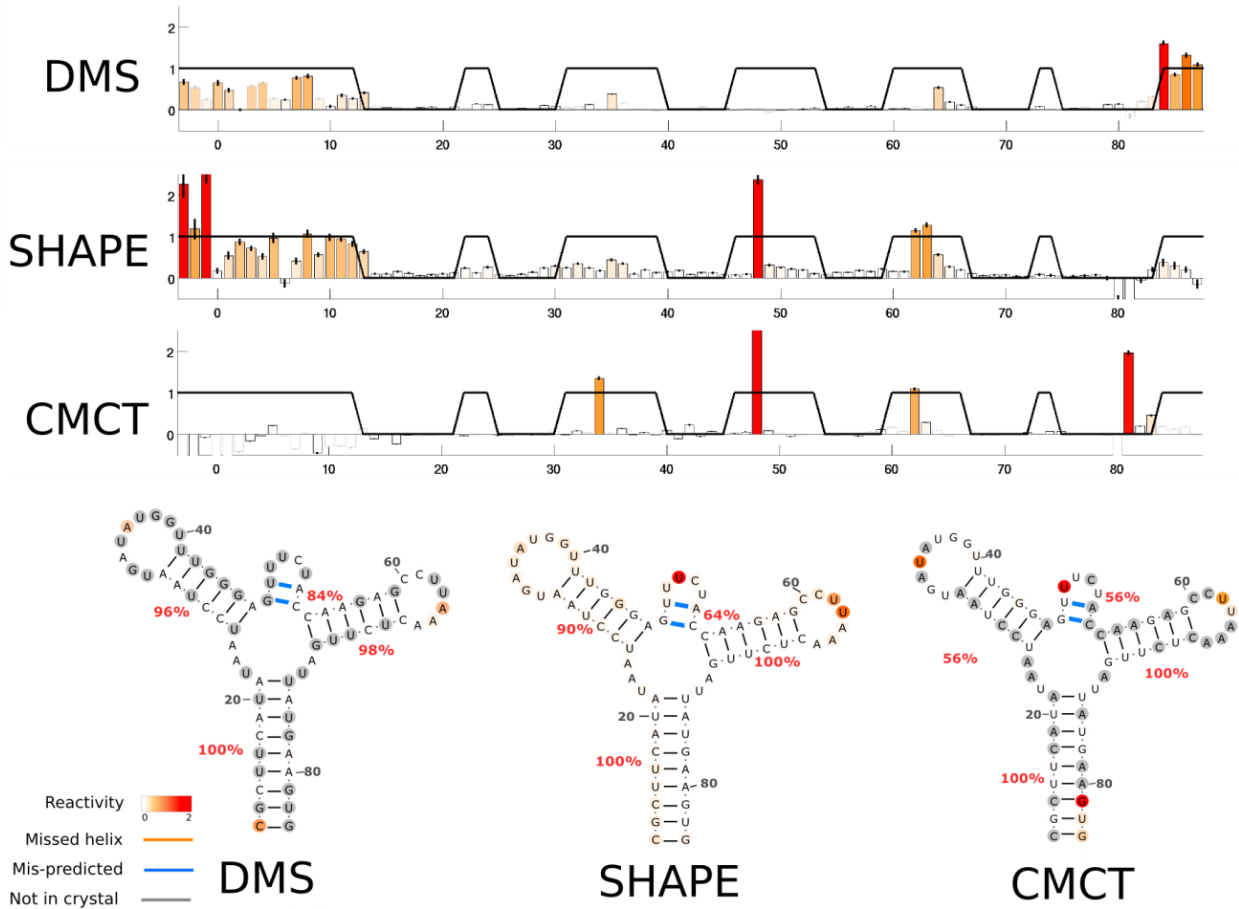

**Figure S5. DMS, SHAPE, and CMCT data and pseudo-energy guided models for tRNA[phe]** – DMS and CMCT data at guanines and uracils, and adenines and cytosines, respectively, are marked in gray.

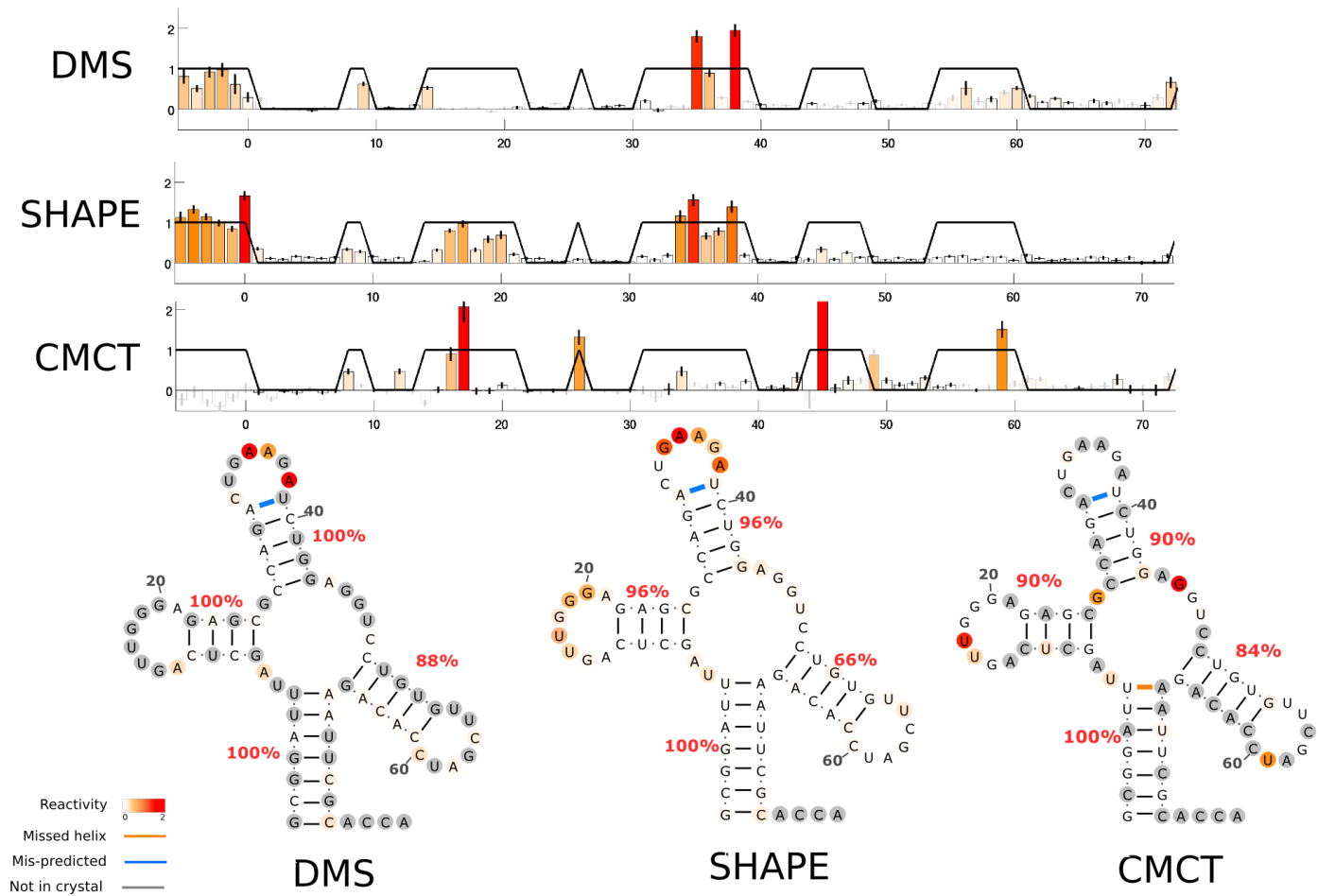

**Figure S6. DMS, SHAPE, and CMCT data and pseudo-energy guided models for cyclic di-GMP riboswitch** – DMS and CMCT data at guanines and uracils, and adenines and cytosines, respectively, are marked in gray.

**Figure S7. DMS, SHAPE, and CMCT data and pseudo-energy guided models for the 5S rRNA** – DMS and CMCT data at guanines and uracils, and adenines and cytosines, respectively, are marked in gray.

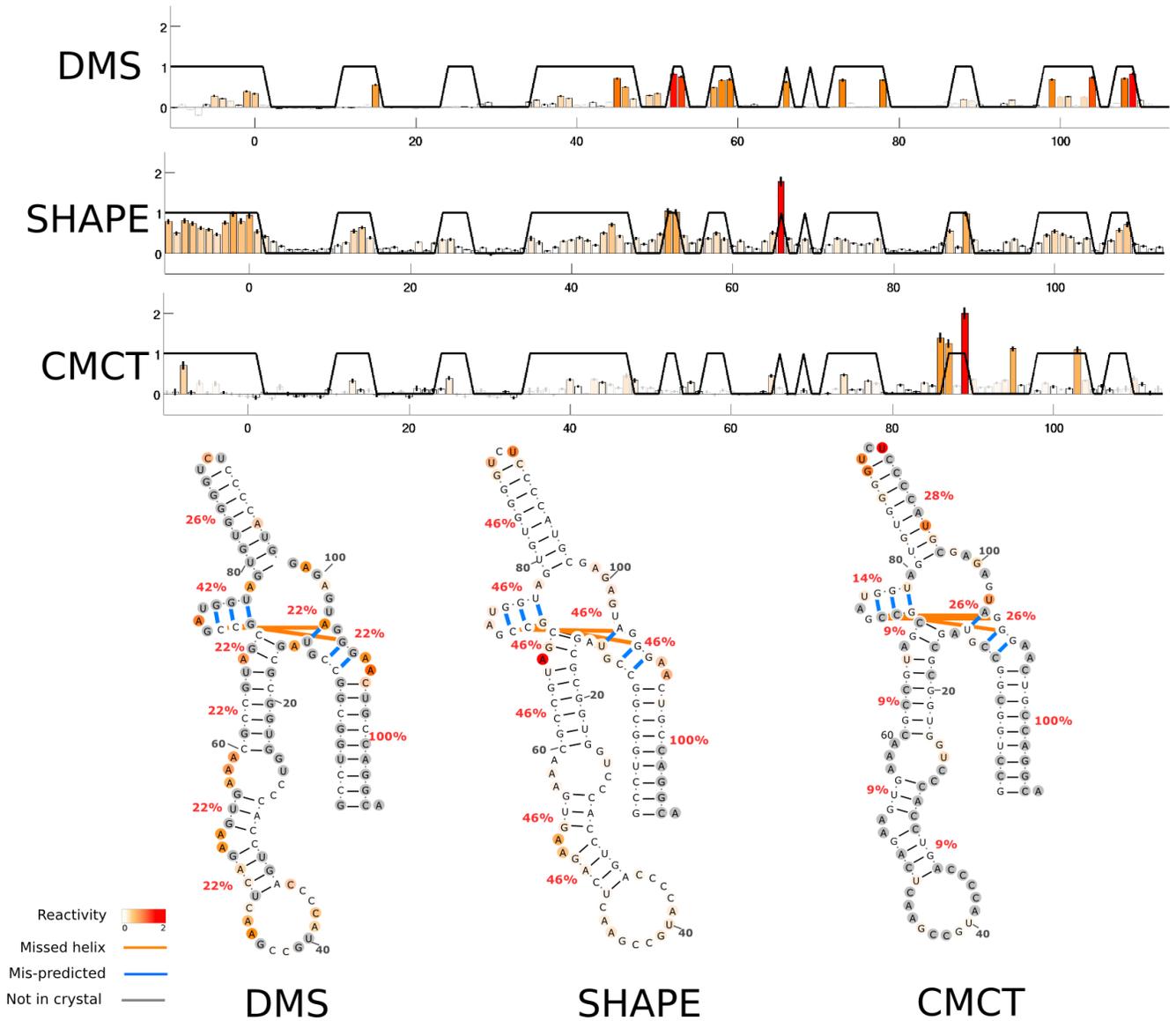

**Figure S8. DMS, SHAPE, and CMCT data and pseudo-energy guided models for the P4-P6 domain of the *Tetrahymena* group I ribozyme–** DMS and CMCT data at guanines and uracils, and adenines and cytosines, respectively, are marked in gray.

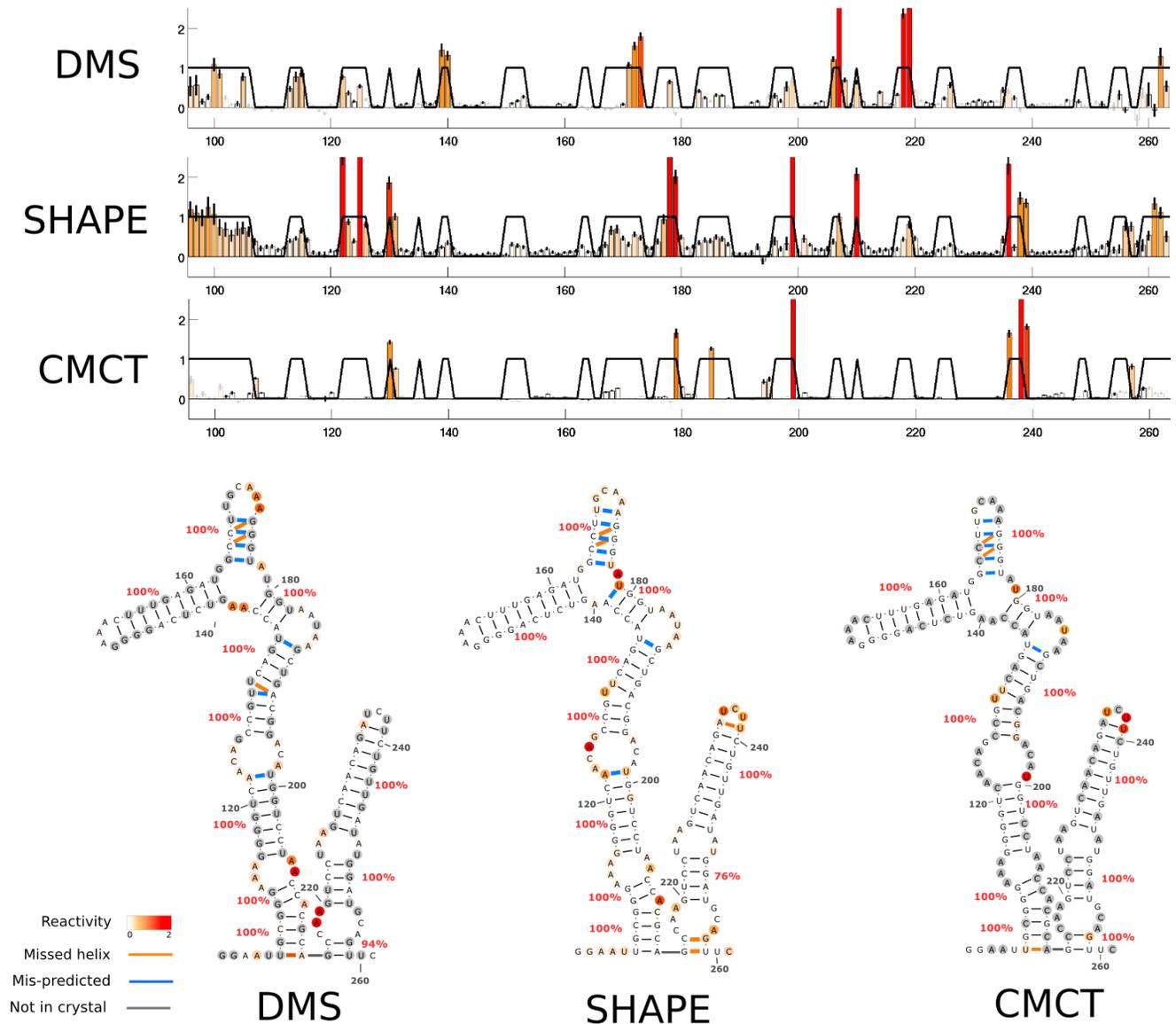

**Figure S9. DMS, SHAPE, and CMCT data and pseudo-energy guided models for the *F. nucleatum* glycine riboswitch–** DMS and CMCT data at guanines and uracils, and adenines and cytosines, respectively, are marked in gray.

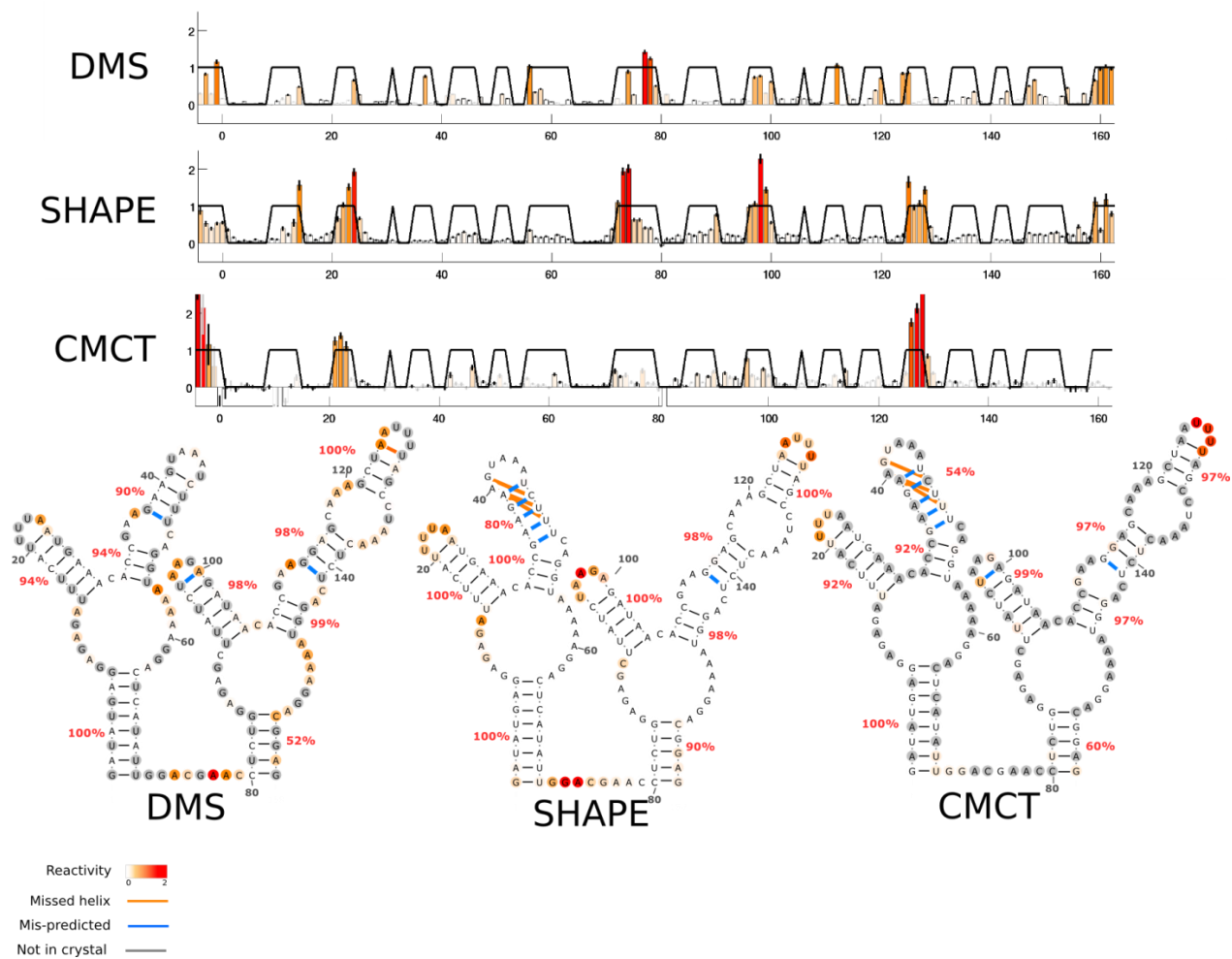

**References for supporting information**

1. Kladwang, W., VanLang, C. C., Cordero, P., and Das, R. (2011) Understanding the errors of SHAPE-directed RNA structure modeling., *Biochemistry*. American Chemical Society *50*, 8049-56.

2. Yoon, S., Kim, J., Hum, J., Kim, H., Park, S., Kladwang, W., and Das, R. (2011) HiTRACE: high-throughput robust analysis for capillary electrophoresis, *Bioinformatics 27*, 1798-1805.

3.  Deigan, K. E., Li, T. W., Mathews, D. H., and Weeks, K. M. (2009) Accurate SHAPE-directed RNA structure determination., *Proceedings of the National Academy of Sciences of the United States of America*. National Academy of Sciences *106*, 97-102.

4.  Rohl, C. A., Strauss, C. E. M., Misura, K. M. S., and Baker, D. (2004) Protein structure prediction using Rosetta., *Methods in enzymology 383*, 66-93.

5.  Rohl, C. A., Strauss, C. E. M., Chivian, D., and Baker, D. (2004) Modeling structurally variable regions in homologous proteins with rosetta., *Proteins 55*, 656-77.

6.  Das, R., and Baker, D. (2007) Automated de novo prediction of native-like RNA tertiary structures., *Proceedings of the National Academy of Sciences of the United States of America 104*, 14664-9.

7.  Kladwang, W., VanLang, C. C., Cordero, P., and Das, R. (2011) A two-dimensional mutate-and-map strategy for non-coding RNA structure., *Nature chemistry*. Nature Publishing Group *3*, 954-62.

8.  Darty, K., Denise, A., and Ponty, Y. (2009) VARNA: Interactive drawing and editing of the RNA secondary structure., *Bioinformatics (Oxford, England) 25*, 1974-5.